\documentclass[11pt,a4paper]{article}
\usepackage{lipsum}
\usepackage{authblk}
\usepackage{graphicx}
\usepackage[format=hang,labelfont=bf,textfont=small,singlelinecheck=false,margin={12pt,12pt},figurename=Fig.]{caption}
\usepackage[top=2cm, bottom=2cm, left=2cm, right=2cm]{geometry}
\usepackage{fancyhdr}
\usepackage{abstract}
\usepackage{hyperref} %for adding email

\begin{document}
%
%title and author details
\title{ Geometry of tilt (in)variance in scanned oblique plane microscopy}
\author[1]{Manish Kumar}
\author[1,*]{Yevgenia Kozorovitskiy}
\affil[1]{Department of Neurobiology, Northwestern University}
\affil[*]{\textit {Yevgenia.Kozorovitskiy@northwestern.edu}}
%\affil[*]{\href{mailto:Yevgenia.Kozorovitskiy@northwestern.edu}{Yevgenia.Kozorovitskiy@northwestern.edu}}
%
\date{}
\maketitle

\begin{abstract}
\noindent Oblique plane microscopy (OPM) is a single objective light-sheet microscopy which performs three dimensional (3D) imaging by axial scan of the generated light-sheet. Recently, multiple techniques for lateral scan of the generated light-sheet in OPM have emerged. However, their suitability for geometrically distortion free 3D imaging, which essentially requires a constant tilt light-sheet scan, has not been evaluated. In this work, we use a geometrical optics approach and derive analytical relationship for the amount of tilt variance in planar mirror based scanned oblique plane illumination (SOPi) arrangement. We experimentally validate the derived relationship and use it to arrive at an optimized scanner geometry and to understand its associated limitations. We also discuss the effects of scanning on optical aberrations and 3D field of view in optimized, tilt invariant, lateral scanning OPM systems.
\end{abstract}

\medskip
\section{Introduction}
\begin{figure}[htbp]
\centering\includegraphics[width=0.9\linewidth]{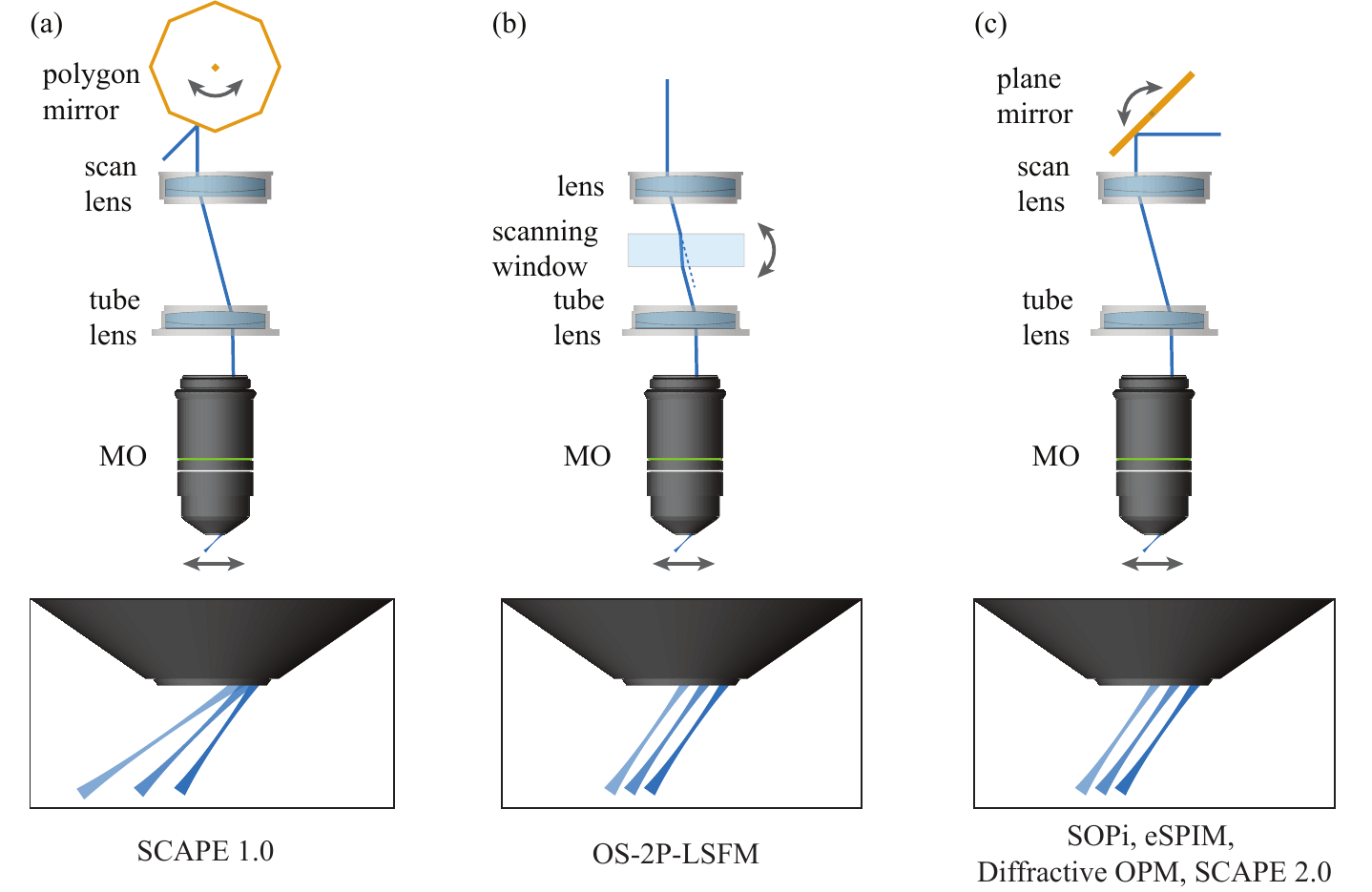}
\caption{Existing geometries for lateral scan of the generated light-sheet in OPM inspired systems. Insets show corresponding light-sheet scan orientations. (a) First arrangement uses a polygon mirror scanner to perform a lateral scan with varying tilt \cite{bouchard2015SCAPE1}. (b) Second arrangement uses a refractive transmission scanning window to perform constant tilt lateral scan \cite{shin2018oblique}. (c) Third arrangement uses a plane mirror scanner to aim for constant tilt lateral scan \cite{kumar2018SOPi1,kumar2019SOPi2,yang2019eSPIM,hoffmann2019diffractive,voleti2019SCAPE2}. MO: microscope objective.}
\label{fig1}
\end{figure}

Light-sheet microscopy is a powerful imaging technique based on optical sectioning. The conventional light-sheet microscopy configuration consists of two objectives orthogonally arranged around a sample \cite{siedentopf1902uber, voie1993orthogonal}. Several variants of light-sheet microscopy have been developed for more convenient sample access during imaging \cite{wu2011inverted,wu2013spatially,holekamp2008fast,tokunaga2008highly,migliori2018light,glaser2019multi}. However, oblique plane microscopy (OPM) is an unique setup which performs light-sheet microscopy with a single objective facing a given sample, thus providing maximum steric access to the sample \cite{dunsby2008OPM1}. To achieve this, it combines the concepts of aberration free remote focusing and selective plane illumination microscopy (SPIM) \cite{botcherby2007aberration,botcherby2008optical,huisken2004optical}. In its original configuration, OPM performs 3D imaging by axial scan of the light-sheet, achieved by piezo mounted remote microscope objective \cite{kumar2011OPM2,sikkel2016high}. Remote objective's axial movement results in an absolutely tilt invariant axial scan of the generated oblique light-sheet, making 3D reconstruction simple and free of geometrical distortions. Recently, multiple alternate scan configurations have been implemented in OPM for a more convenient, lateral scan of the generated light-sheet (see Fig. \ref{fig1}). Swept confocally-aligned planar excitation (SCAPE) microscopy was the first among these to introduce a polygon scan mirror based reflective arrangement, leading to lateral scan of the light-sheet \cite{bouchard2015SCAPE1}. A second configuration demonstrated oblique scanning two-photon light-sheet fluorescence microscopy (OS-2P-LSFM), which made use of a refractive transmission window for lateral scan of the light-sheet \cite{shin2018oblique}. However, these scan arrangements were associated with several limitations. SCAPE introduced a scan position dependent tilt variation in the light-sheet, which is suboptimal for geometrical distortion free 3D imaging. The refractive window solution provided an absolutely tilt invariant scan for the excitation light-sheet, but not for the imaging path. Since this type of imaging relies on a relatively wide range (wavelength) of fluorescence signals, optical dispersion and aberrations (primarily spherical) are generated because of high refractive index glass window in the imaging path.

To combine the best features of the previous two arrangements, we introduced a plane mirror based scanned oblique plane illumination (SOPi) microscopy \cite{kumar2018SOPi1, kumar2019SOPi2}. Our work focused on optical ray tracing simulations to obtain an optimized scan geometry to resolve the tilt variance problem for both scanned illumination and descanned signal rays. Here, a plane mirror scanner is placed with its rotation axis at the intersection of the back focal plane (BFP) and the principal axis of the scan lens. In parallel, Yang \textit{et al.} independently introduced epi-illumination SPIM (eSPIM) with a plane mirror scanner for lateral scanning light-sheet \cite{yang2019eSPIM}. However, eSPIM focused on solving the low effective numerical aperture issue of the OPM systems, and it did not delve into the scanner geometry.  Subsequently, there has been a steady rise in the popularity of plane mirror scan geometry for creating systems with direct application in developmental biology and neuroscience. Two notable implementations include diffractive OPM and SCAPE 2.0 \cite{hoffmann2019diffractive,voleti2019SCAPE2}. Diffractive OPM performs single objective light-sheet imaging with low numerical aperture objectives and results in a very large field of view imaging in small organisms. SCAPE 2.0 demonstrates the rapid imaging capability of lateral scan architecture in OPM by imaging unrestricted small organisms at cellular resolution. 

Given the challenges of the first two scan geometries in Fig. \ref{fig1}, plane mirror scan geometry is poised to become the preferred arrangement in future developments in this family of imaging techniques. However, plane mirror scanner geometry has not yet been studied in sufficient detail, in order to understand the underlying principles and, most importantly, the limitations of this scan geometry in OPM. Here, we perform a detailed geometrical analysis of this scan arrangement. We derive a relationship for evaluating tilt variance in scanned light-sheet. We also perform an experiment to measure the actual variation in the tilt of an oblique beam and cross validate the derived relationship. We then use the derived relationship to arrive at an optimized scanner placement. In addition to addressing tilt invariance in oblique light-sheet scanning, we also evaluate optical aberrations and field of view in the optimized system during a lateral scan. 

%%%%%%%%%%%%%%%%%%%%%%%%%%  Geometrical Analysis  %%%%%%%%%%%%%%%%%%%%%%%%%%
\section{Geometrical investigations of tilt invariant lateral scan and imaging}
Geometrical optics is a simple yet powerful tool for analyzing optical systems. These intuitive analyses describe light as optical rays which travel in straight lines, bending or reflecting at interfaces, governed by well-known laws of refraction and reflection. Geometrical analysis is capable of investigating imaging performance and aberrations in an optical system \cite{born2013principles}. Various ray tracing software, which are industry and research standards for optical system design and optimization, rely on concepts from geometrical optics \cite{fischer2008optical}.  In our earlier work, we performed ray tracing based optimization for arriving at the SOPi arrangement \cite{kumar2018SOPi1}. It remains unknown whether the scan is absolutely tilt invariant, and if not, its deviation from expected ideal behavior. Therefore, in this section we build a thorough geometrical analysis of the plane mirror based scanner, aimed towards tilt invariant scan through an optical lens. We analyze the behavior of optical rays in a single plane first, as their extension into a light-sheet configuration is straightforward. 

\begin{figure}[htbp]
\centering\includegraphics[width=0.9\linewidth]{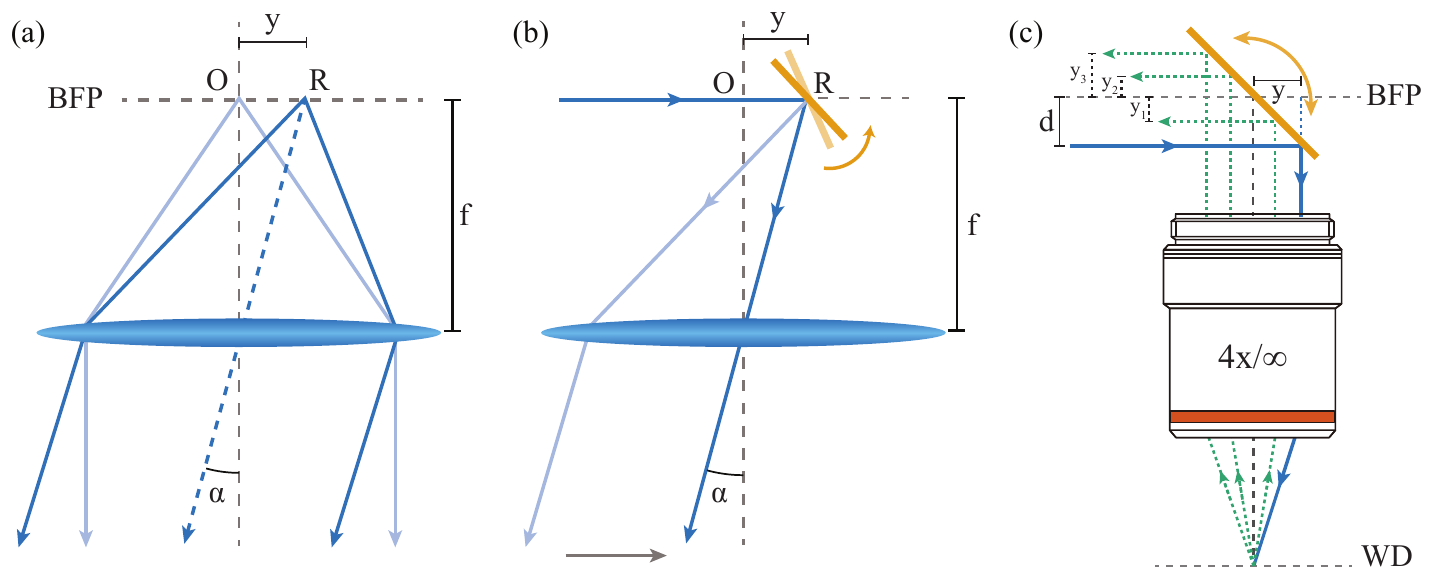}
\caption{Lens as an optical Fourier transforming element for tilt invariant scan and imaging. (a) A point source at the BFP leads to collimated optical rays where the lateral offset location of point source determines tilt angle of collimated rays. (b) A mirror scanner centered at the BFP leads to an absolutely tilt invariant lateral scan. (c) SOPi scan geometry under consideration for tilt invariant scan and imaging. WD: working distance.}
\label{fig2}
\end{figure}

\subsection{Ideal lens, Fourier transform, and tilt invariant lateral scan}
An optical lens is well known to behave as a Fourier transforming element \cite{goodman2005introduction}. A point source placed on the BFP of an ideal thin lens (optical aberration free) provides a set of collimated optical rays. As shown in Fig. \ref{fig2}(a), a point source offset by distance $y$ leads to collimated optical rays with tilt angle $\alpha$.  In other words, if an optical ray emerges from a lens with a known tilt $\alpha$, it can be uniquely associated with an offset point $R$ on the BFP which is $y = f \times tan(\alpha)$ distance apart from the principal axis, where $f$ is the focal length of the lens. This property of an optical lens can be built upon to describe an ideal tilt invariant scan. Let us consider a planar mirror scanner placed with its rotation axis at $R$ (on the BFP)  as shown in Fig. \ref{fig2}(b). A laser beam hits the scanner at $R$, to get reflected towards the lens. Since the pivot point of reflected ray is fixed at $R$ (on the BFP), it leads to a constant tilt lateral scan beyond the lens. This tilt angle
\begin{equation}
\alpha = tan^{-1}\left(\frac{y}{f}\right),
\label{eq:eqn1}
\end{equation}
can be easily changed by shifting the scanner and hence the optical beam pivot point $R$ along the BFP. 

\subsection{Tilt invariant lateral scan and imaging}
The geometry shown in Fig. \ref{fig2}(b) provides an absolutely tilt invariant scan of the oblique illumination beam. However, OPM is not limited to the consideration of excitation beam alone. It also requires consideration of the signal rays, arising due to optical scattering or emitted fluorescence from the sample. Unlike excitation beam, signal rays in OPM are not confined to a particular tilt angle. Therefore, we need an optical scanner which provides tilt invariant scanning/descanning for a wide range of beam offsets. Figure \ref{fig2}(c) shows the SOPi arrangement under consideration for this task. An infinity corrected microscope objective serves as a Fourier transforming scan lens, and a plane mirror with its rotation axis at the intersection of the BFP and the principal axis operates as a scanner. A beam (blue line) with offset $y$ forms an oblique illumination beam, and signal optical rays (green dotted) emerge at various tilt angles, where each tilt angle corresponds to a unique offset value $y_1, y_2, y_3,$ etc. What remains to be determined is how the beam offset and therefore tilt angle of the optical rays change during scanning. Tilt variance in oblique optical beam would lead to distorted 3D scan of the sample, while beam offset dependent tilt variance would cause additional optical aberrations \cite{fischer2008optical}.

\subsection{Geometrical derivation}
\begin{figure}[htbp]
\centering\includegraphics[width=\linewidth]{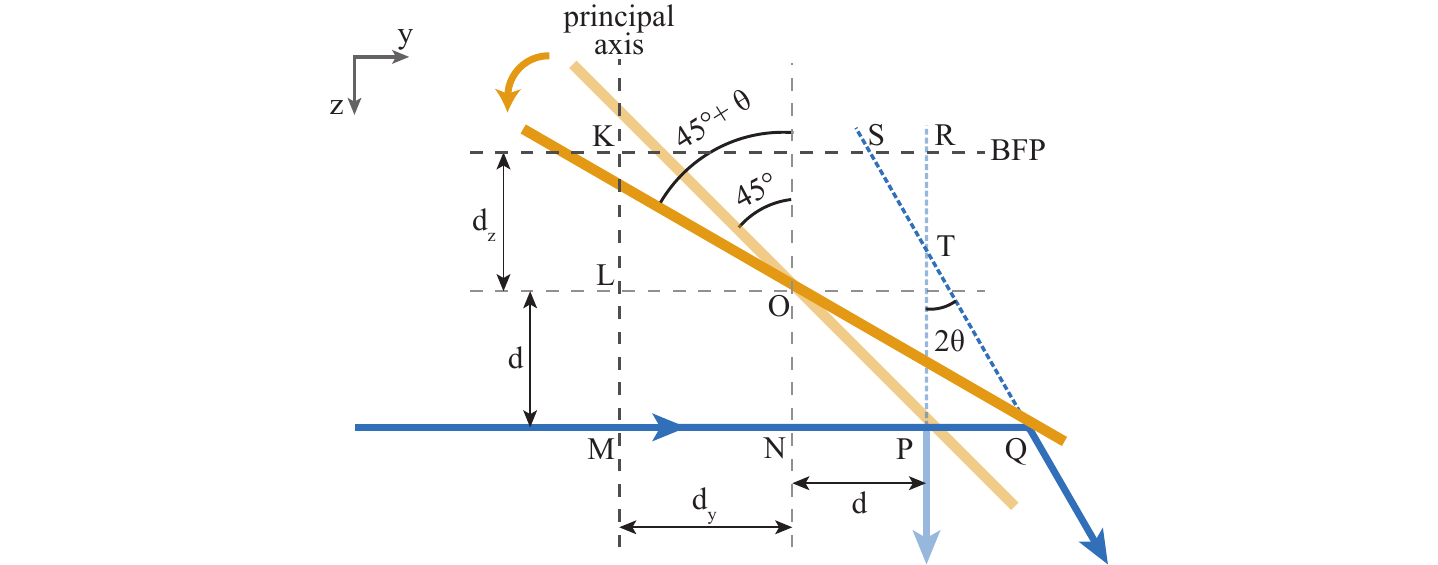}
\caption{A generalized SOPi scan geometry for the evaluation of tilt variance.}
\label{fig3}
\end{figure}

In this section we derive the relationship for scan dynamics of an optical beam in the SOPi geometry (Fig. \ref{fig2}(c)). For a generalized approach, we consider a scan geometry where the rotation axis of the scanner $O$ is offset by $d_y$ and $d_z$ lengths along $y$ and $z$ axis, respectively. Figure \ref{fig3} shows the magnified geometrical optics picture of this arrangement. $OL$ and $KL$ represent the horizontal and vertical offsets of scanner rotation axis from the intersection point of the principal axis and the BFP, respectively. Thus, $OL = d_y$ and $KL = d_z$. An optical ray $MN$ is incident parallel to the BFP with an offset $d$ from the scanner rotation axis $O$. This ray crosses the principal axis at $M$ and hits the $45^{\circ}$ tilted scan mirror (light orange) at $P$ to get reflected vertically downwards along the $z$ axis. When extended, the reflected beam meets the BFP at $R$. Thus, $ON\perp NP$, $NP\perp PR$, $KM = PR = d+d_z$, and $ON = LM = NP = d$. We now consider a new scanner position (dark orange) with the tilt angle $45^{\circ} + \theta$. Following the laws of reflection, the optical ray now hits the scanner at $Q$ and is reflected, making an angle $2\theta$ with the $z$ axis. This reflected optical ray, when traced backwards, meets the BFP at $S$. $T$ is the intersection point of both reflected rays where $\angle PTQ = \angle RTS = 2\theta$. For an ideal scanner geometry, $R$ and $S$ should overlap, leading to a constant offset and hence an absolutely tilt invariant scan. However, in practice, the gap $RS$ dictates the error, or tilt variance, during the scan.

In $\Delta ONQ$ we have $tan(\angle NOQ) = NQ/ON = (NP+PQ)/ON = 1 + PQ/d$. 
Therefore, $PQ = d\times(tan(\angle NOQ) -1) = d\times(tan(45^{\circ}+\theta)-1) = d\times[(1+tan\theta)/(1-tan\theta)-1]$.
\\Or, \begin{equation}
PQ = \frac{2d \times tan\theta}{1-tan\theta} .
\label{eq:eqn2}
\end{equation}
In $\Delta TPQ$ we have $tan(\angle PTQ) = PQ/PT = PQ/(PR-RT) = PQ/(d+d_z-RT)$.
Therefore, $RT = d + d_z - PQ/tan(\angle PTQ) = d + d_z - PQ/tan(2\theta)$. Replacing $PQ$ from Eq. \ref{eq:eqn2} and expanding $tan(2\theta)$ we get $RT = d + d_z - d\times (1+tan\theta)$. Or, 
\begin{equation}
RT = d_z - d \times tan\theta .
\label{eq:eqn3}
\end{equation}
In $\Delta TRS$ we have $RS = RT\times tan(\angle RTS)$. Or,
\begin{equation}
\label{eq:eqn4}
RS = (d_z - d \times tan\theta) \times tan(2\theta),
\end{equation}
where we replaced $RT$ from Eq. \ref{eq:eqn3}. Here, we can use $RS$ to precisely calculate tilt variance in the oblique optical beam during scan. The practical value of scan angle is $\theta <5^{\circ}$, implying that $RS$ is smaller than $d_z$.  If $f$ is the focal length of the Fourier transforming lens, we can use Eq. \ref{eq:eqn1} to express tilt variance in an optical beam as a difference between beam tilt for two values of the mirror tilt \textit{i.e.},  $\delta = [\alpha_0 - \alpha_\theta] = [tan^{-1}(KR/f) - tan^{-1}(KS/f)]$, where $\alpha_0$ and $\alpha_\theta$ correspond to $45^\circ$ and $(45^\circ+\theta)$ mirror tilt angles, respectively. Rewriting $KR$ and $KS$ we have
\begin{equation}
\label{eq:eqn5}
\delta = \left[tan^{-1}\left(\frac{d+d_y}{f}\right) - tan^{-1}\left(\frac{d+d_y-RS}{f}\right)\right]. 
\end{equation}
In a practical case, Fourier transforming lenses have much longer focal lengths, compared to beam offsets and scanner position offsets (\textit{e.g.} ref. \cite{kumar2019SOPi2} used $f = 100$ mm and offset $d = 3.54$ mm). Therefore, we have $(d+d_y)/f \ll 1$ and $(d+d_y-RS)/f \ll 1$ leading to the following small angle approximation $\delta \approx (d+d_y)/f - (d+d_y-RS)/f = RS/f$. Or, replacing $RS$ from Eq. \ref{eq:eqn4} we have
\begin{equation}
\label{eq:eqn6}
\delta \approx \left[ \frac{d_z \times tan(2\theta)}{f} - \frac{d\times tan(\theta)\times tan(2\theta)}{f} \right],
\end{equation}
where $\delta$ is in radians. 

Several considerations follow from Eq. \ref{eq:eqn6}. For a given practical value of tilt angle ($\theta<5^{\circ}$), the first term of the equation is at least one order of magnitude larger than the second term. This implies that $d_z$ plays a greater role in tilt variance. On the other hand, the second term is responsible for beam offset dependent tilt variance, and it may lead to optical aberrations in the system. Notably, tilt variance of the system increases with scan angle $\theta$. On a closer inspection of Eq. \ref{eq:eqn6} (and Eq. \ref{eq:eqn4}), it becomes clear that $d_z = (d\times tan\theta)$ makes $RS$ zero, leading to an absolutely tilt invariant scan. However, this relationship cannot be satisfied for a wide range of $\theta$ unless $d_z = d = 0$. This happens when both the incident beam and the rotation axis of the scanner are aligned to the BFP, \textit{i.e.} the ideal scan condition as shown in Fig. \ref{fig2}(b). If $\theta$ is restricted to small values, a nonzero $d$ is allowed when $d_z$ approaches zero. This optimized case matches the schematic shown in Fig. \ref{fig2}(c), and is consistent with the previously published geometry of SOPi \cite{kumar2018SOPi1,kumar2019SOPi2}. We can further conclude from the expression for $RS$ that an offset along lateral direction ($d_y$) does not change tilt variance in the system. However, a non-zero $d_y$ would change overall tilt of the oblique beam (see Eq. \ref{eq:eqn1}) and an off-axis placement of the scanner would make the imaging system asymmetric. Therefore, the optimal scanning arrangement is one with $d_y = d_z = 0$. Here, the tilt variance expression from Eq. \ref{eq:eqn5} becomes 
\begin{equation}
\label{eq:eqn7}
\delta = \left[tan^{-1}\left(\frac{d}{f}\right) - tan^{-1}\left(\frac{d-RS}{f}\right)\right], 
\end{equation}
which under large focal length and small offsets approximation reduces to $\delta \approx RS/f = -d/f \times tan(\theta)\times tan(2\theta)$. Note that $\delta$ is in radians here. Considering an extreme example with a large scan angle $\theta=10^{\circ}$, a large offset $d=10$ mm, and $f=100$ mm, we get tilt variance $\delta \approx 0.37^{\circ}$. This value of tilt variance is small for most practical purposes. However, based on subsequent optical elements, tilt variance can get magnified to become substantial. For example, the SOPi setup in ref. \cite{kumar2018SOPi1} introduces $22.22 \times$ angular magnification, leading to effective $\sim 8.2^{\circ}$ tilt variance in the sample volume for the theoretical case described above.

\section{Experimental validation}
\begin{figure}[htbp]
\centering\includegraphics[width=\linewidth]{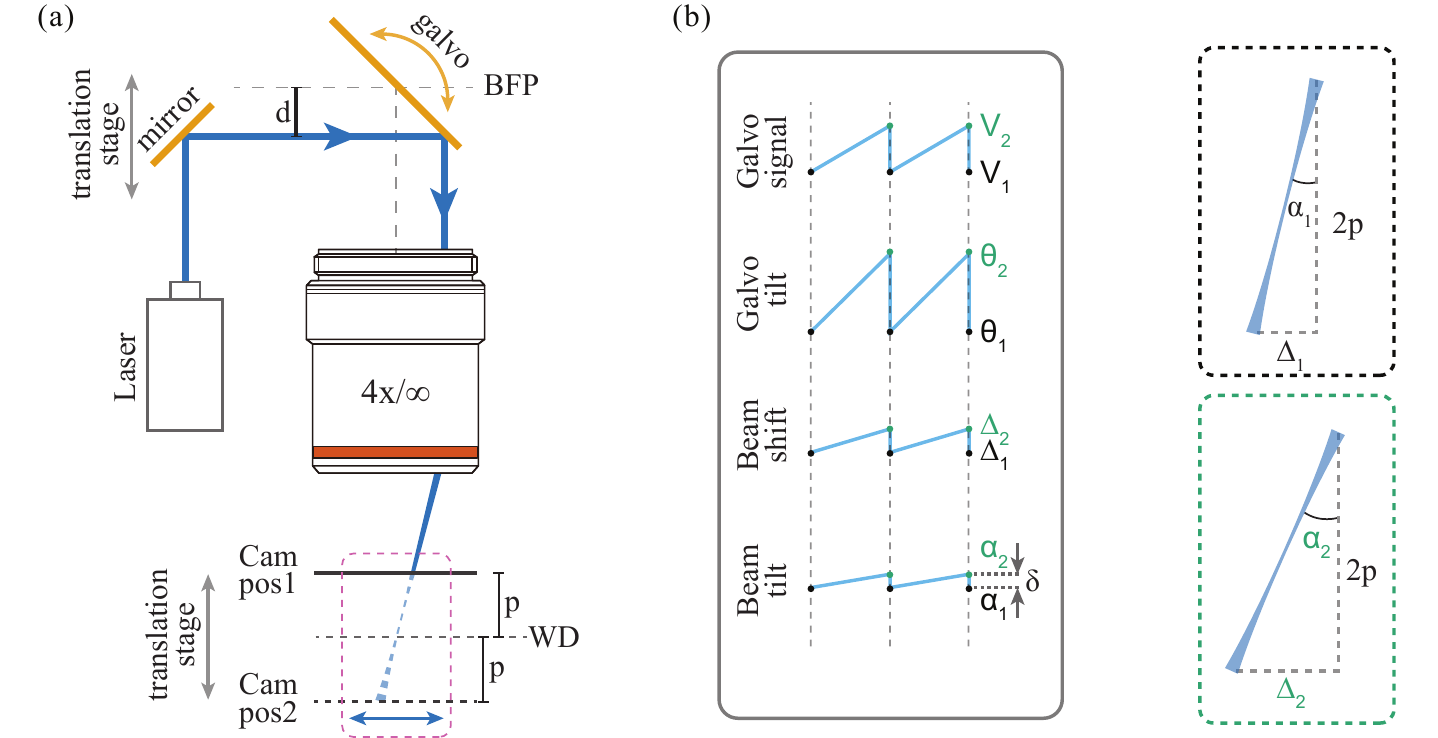}
\caption{Experimental strategy for evaluation of tilt variance. (a) Schematic of the experimental setup for measuring tilt variance. (b) Dependencies between voltage applied to galvo, galvo tilt, and beam shift/tilt.}
\label{fig4}
\end{figure}

Next, we performed an experimental validation of the derived tilt variance relationship. For this we needed a Fourier transforming lens, a plane mirror based scanner, and a method for measuring the beam tilt angle $\alpha$. The schematics of the setup are shown in Fig. \ref{fig3}. We used a low magnification, long working distance microscope objective ($4\times$, 0.1 NA, f = 50 mm, WD = 30 mm, Nikon) as a Fourier transforming lens. The advantage of using a low magnification objective is that its BFP lies outside the body of the objective and is therefore directly accessible without the need for a pair of lenses to relay it to the scanner. A galvanometer mounted plane mirror (QS12, 10 mm aperture, Nutfield) served as the scanner. Since the precise placement of the galvo scanner was crucial, we first directed a collimated laser beam backwards through the microscope objective. This beam converged at the BFP of the objective, where the galvo scanner was carefully aligned to match the convergence point at the BFP with its rotation axis. We used a HeNe laser (HNL100L, Thorlabs) for alignment and experiment. We used a neutral density filter (not shown in figure) to reduce the laser power and reflected the laser beam towards the galvo scanner using a mirror mounted on a manual translation stage. This precision translation stage helped in varying the offset $d$ for the incident laser beam. We used a precision translation stage mounted camera as a tool for measuring the outgoing beam tilt $\alpha$. As shown in Fig. \ref{fig4}(a), the camera sensor plane was oriented perpendicular to the principal axis, and it served to capture beam position at two predefined positions $\pm p$ distance away from the working distance of the microscope objective. This arrangement enabled the calculation of beam tilt 
\begin{equation}
\label{eq:eqn8}
\alpha = tan^{-1}\left( \frac{\Delta}{2p} \right),
\end{equation}
where $\Delta$ is the absolute shift between the beam positions on the two planes (see inset Fig. \ref{fig3}). We used a camera with 3.45 $\mu m$ pixel size (Blackfly, BFS-U3-16S2M-CS, FLIR) and $p = 10$ mm for our experiments. This arrangement has beam tilt measurement resolution of $\sim 0.01^{\circ}$. Since beam divergence led to large beam size on camera, the center of each circular spot was noted as the beam position. As depicted in Fig. \ref{fig4}(b), the beam tilt measurements at two extreme scan points corresponding to galvanometer tilts $\theta_1$ (for $V_1$ voltage) and $\theta_2$ (for $V_2$ voltage) determined the tilt variance  
\begin{equation}
\label{eq:eqn9}
\delta_{exp}=\alpha_2 - \alpha_1 = tan^{-1}\left( \frac{\Delta_2}{2p} \right) - tan^{-1}\left( \frac{\Delta_1}{2p} \right),
\end{equation}
where $\Delta \ll 2p$ invokes small angle approximation leading to $\delta_{exp} = (\Delta_2-\Delta_1)/2p$ (in radians). During experiments, a third precision translation stage (not shown in figure) helped shift the microscope objective and hence control $d_z$. Even with careful alignment of the camera linear translation stage, a slight angular mismatch in its translation axis and the microscope objective's principal axis is unavoidable. This mismatch gives rise to a consistent offset in the beam position, when measured at two camera positions. This offset is a constant value that can be easily compensated for by recording the on-axis beam positions (with $d=0$ mm) at the two planes.

We performed experiments for $d_z = 0$ mm and $d_z = 8$ mm. For each of these cases, we recorded the beam positions on two planes ($2p=20$ mm apart), with two voltages $V_{1,2} = \pm 0.4$, and four offset values $d =$ 0 mm, 1 mm, 2 mm, and 3 mm. For each of these combinations, we obtained $\Delta_1$ and $\Delta_2$ as displayed in the unit of pixels (px) in Table \ref{table1}. We experimentally measured the galvanometer's tilt angle $\theta$ as $\pm 0.92^{\circ}$, in response to the applied $\pm 0.4$ V. This was determined by measuring the deflection in galvanometer reflected laser beam, propagating through air onto a distal screen, in response to the applied voltage and using $\theta = 0.5 \times tan^{-1}(deflection \div screen \; distance) = 0.5 \times tan^{-1}(\pm 2.4 \, mm/74.5 \, mm)$. We confirmed that the small angle approximation is valid with our choice of parameters. For example, the experimental case with $d_z = 8$ mm, $d = 3$ mm, $f = 50$ mm and $\theta = \pm 0.92^{\circ}$ yields very similar values for $\delta_{th}$ as $0.587^{\circ}$ (using Eq. \ref{eq:eqn5}), and $0.589^{\circ}$ (using Eq. \ref{eq:eqn6}). We used Eq. \ref{eq:eqn6} to calculate $\delta_{th}$ values in Table \ref{table1}. We filled in $\Delta_1$ and $\Delta_2$ values from the experimental measurements of beam positions and used Eq. \ref{eq:eqn9} to calculate $\delta_{exp}$.

\begin{table*}[t]
\centering
\caption{\bf Calculation of Tilt Variance During Scan}
\begin{tabular}{cccccc}
\hline
 $d_z$ (mm) & $d$ (mm) & $\delta_{th}$ & $\Delta_1$ (px) & $\Delta_2$ (px) & $\delta_{exp}$\\
\hline
0 & 0 & $0.00^{\circ}$ & 0 & -5 & $-0.05^{\circ}$ \\
0 & 1 & $0.00^{\circ}$ & 170 & 164 & $-0.06^{\circ}$ \\
0 & 2 & $0.00^{\circ}$ & 244 & 237 & $-0.07^{\circ}$ \\
0 & 3 & $0.00^{\circ}$ & 367 & 360 & $-0.07^{\circ}$ \\
%\hline
8 & 0 & $0.59^{\circ}$ & -28 & 25 & $0.53^{\circ}$ \\
8 & 1 & $0.59^{\circ}$ & 88 & 142 & $0.54^{\circ}$ \\
8 & 2 & $0.59^{\circ}$ & 201 & 255 & $0.54^{\circ}$ \\
8 & 3 & $0.59^{\circ}$ & 315 & 370 & $0.55^{\circ}$ \\
\hline
\end{tabular}
  \label{table1}
\end{table*}

It is evident from Table \ref{table1} that the theoretical and experimental values of tilt variances tightly match. Moreover, it is clear that tilt variance is mainly dependent on $d_z$ and has negligible dependence on beam offset $d$. A closer inspection shows that $\delta_{exp}$ is consistently offset along one direction from $\delta_{th}$ (average offset = $-0.06^{\circ}$). This consistent offset can be explained by an unintentional residual $d_z$ remaining in the setup during galvanometer alignment. In fact, Eq. \ref{eq:eqn6} translates $-0.06^{\circ}$ offset in $\delta$ into $-800 \, \mu m$ offset in $d_z$. Therefore, galvanometer position can be compensated by this length to obtain a perfect alignment for tilt invariant scan. This analysis highlights the following aspects of SOPi microscopy. First, the SOPi system is highly sensitive to galvanometer positioning. Second, even careful experimental alignment may not be precise enough to obtain tilt invariant scan. Third, measurement of tilt variance during scan, combined with our derived relationships, can be used for precise measurement and correction of galvanometer position. Fourth, tilt variance for the corrected system can approach zero, leading to a practically tilt invariant scan. 

\section{Optical aberrations and field of view during SOPi scan}
We have demonstrated that the SOPi scan geometry can offer tilt invariant scanning and imaging with an oblique light-sheet. However, it remains important to consider any limitations of this scan geometry. In this section we assess the effects of scanning on optical aberrations and field of view. We continue with a geometrical optics approach for these analyses.

\subsection{Optical aberrations during scanning} 
\begin{figure}[htbp]
\centering\includegraphics[width=\linewidth]{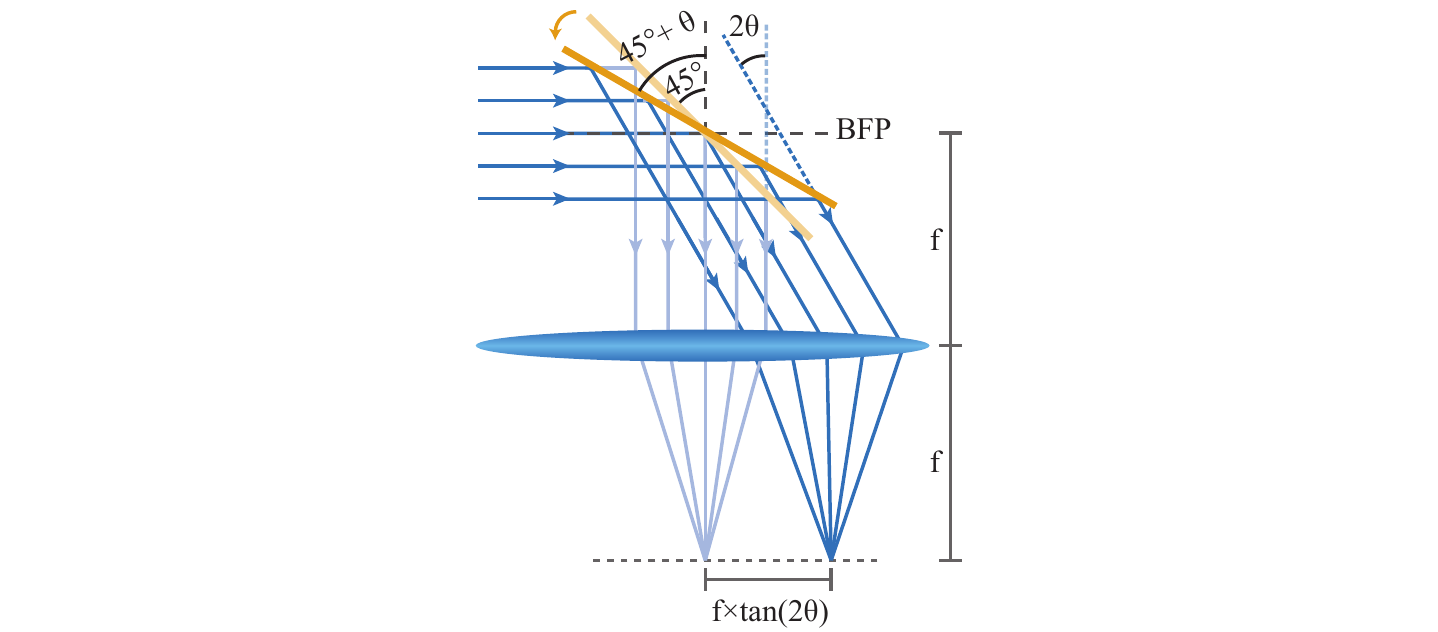}
\caption{Effect of lateral scanning on optical aberrations.}
\label{fig5}
\end{figure}
\noindent Optical aberrations in a lens can be evaluated by tracing a pencil of parallel optical rays and observing how well they converge \cite{mahajan1991aberration,fischer2008optical}. An unaberrated optical lens leads all of the parallel rays to converge to a single point. Any deviation from this behavior is credited to the presence of optical aberrations. To evaluate optical aberrations due to the previously described scan arrangement, we consider an optical aberrations free lens in the optimized scan geometry where $d_z=0$. This leads to $RS = -  d\times tan\theta \times tan(2\theta)$ (from Eq. \ref{eq:eqn4}), and corresponding tilt variance $\delta = RS/f = -  d/f\times tan\theta \times tan(2\theta)$ (from Eq. \ref{eq:eqn6}). There remains a residual tilt variance term which is proportional to $d$. In other words, an optical ray undergoes different amount of tilt variance based on its offset position from the principal axis. At first, it appears that this beam offset dependent tilt variance would lead to optical aberrations in the system. However, this is not the case. A pencil of optical rays get stretched (or compressed) based on the offset of individual rays. This stretching causes a change in tilt of the rays beyond the lens. Figure \ref{fig5} shows that all optical rays, following the law of reflection, make a constant $2\theta$ angle with the principal axis. This angle is independent of offset $d$. Therefore, these rays perfectly converge to a point on the focal plane, $f\times tan(2\theta)$ away from the principal axis. It is important to note that we have considered an ideal optical lens. A real optical lens may show some deviation from these ideal characteristics due to the presence of optical aberrations within the lens. However, the plane mirror scanner arrangement does not add any optical aberrations.

\subsection{Three dimensional field of view during scan}
\begin{figure}[htbp]
\centering\includegraphics[width=\linewidth]{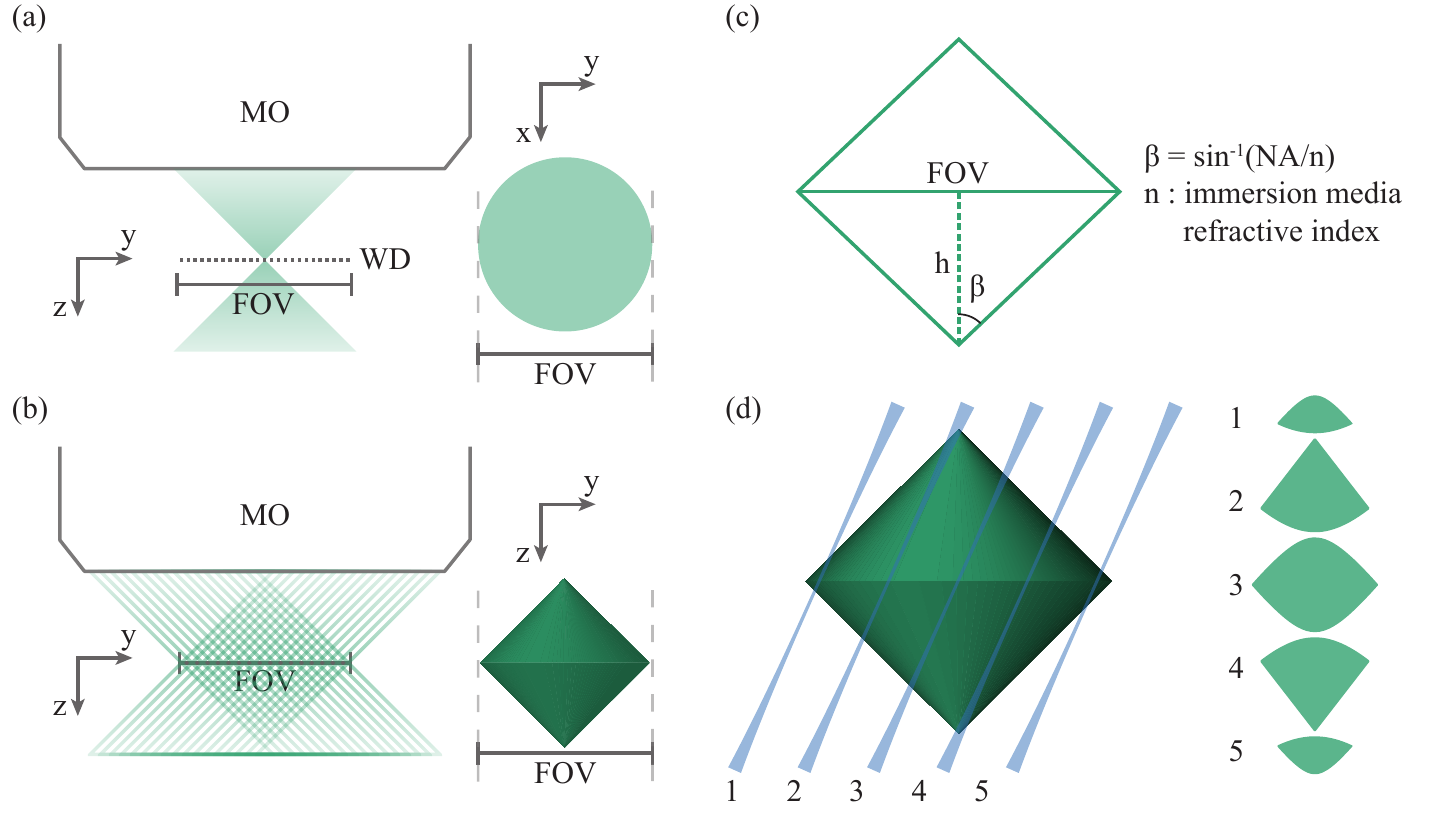}
\caption{3D field of SOPi-like microscopy. (a) Acceptance cone of a microscope objective and corresponding two dimensional field of view. (b) A complete picture of acceptance cones through two dimensional field of view, where the overlapping region (double-cone shape) defines the three dimensional field of view. (c) Relationship between 2D field of view, numerical aperture, and  3D field of view. (d) Light-sheet orientation and corresponding cross-sectional field of view during lateral scan in SOPi microscopy.}
\label{fig6}
\end{figure}

So far we have seen that SOPi arrangement provides tilt invariant lateral scan and adds no optical aberrations in the system. Next, we consider the field of view (FOV) characteristics of SOPi arrangement. A microscope objective is designed for a particular FOV which, as illustrated in Fig. \ref{fig6}(a), is specified as a disk of certain diameter at the working distance of the objective. A point lying outside the FOV is not imaged sharply, due to clipping of a subset of optical rays, \textit{i.e.} vignetting. SOPi and related systems perform 3D imaging, requiring a consideration of the 3D FOV. Moreover, even the 2D FOV of the system is unusual due to the oblique nature of the light-sheet, requiring careful consideration.

Figure \ref{fig6}(b) shows the set of all acceptance cones through the 2D FOV of the microscope objective. For illustration purposes, we have shown a cross section view with the edges of each cone and have made them equidistant within the region. Clearly, any point lying outside the crossed lines region will not get completely covered by the total acceptance cone angle of the objective. Therefore, 3D FOV of the system is essentially a double cone shaped region as shown on the right in Fig. \ref{fig6}(b). In Fig. \ref{fig6}(c) we see that the double cone shape is made of two identical cones of height $FOV\div (2\times tan\beta)$ joined at their bases. Here, $\beta = sin^{-1}(NA/n)$ is the half acceptance angle of the imaging system. All points inside this double cone 3D FOV are imaged sharply by a SOPi-like microscope. At a given scan position, 2D FOV would be represented by the intersection of 3D FOV and the light-sheet plane. Thus, 2D FOV in SOPi microscopy varies with lateral scan position and tilt angle of the light-sheet, potentially limiting the overall system lateral scan range. Figure \ref{fig6}(d) shows the effective 2D FOV of a SOPi-like microscopy at various lateral scan positions. It is apparent that the lateral scan range is large for thin samples and reduces with an increase in sample thickness ($z$-axis range). 

Note that within this 3D FOV, the effective NA of the system depends on selection of the optical elements. SOPi's effective NA can be calculated using \textit{Crossbill Design}, a Python based, platform independent, user friendly GUI for designing oblique light-sheet microscopes  \cite{CrossbillDesign}. 

\section{Conclusion}
We have performed a detailed geometrical analysis of tilt variance in scanned oblique plane microscopy, defined the optimal layout analytically, developed an experimental method and performed a measurement of tilt variance in a specific objective and scanner arrangement. These results confirm that essentially tilt invariant scanning can be achieved by lateral scan implementations of OPM inspired systems, but highlight the importance of precise scanner positioning and alignment for tilt variance and imaging performance control. Moreover, the experimental measurement of tilt variance, combined with our derived analytical relationship, can be used as a tool for precision alignment and positioning of scanners in these systems. We have also pointed out the absence of additional optical aberrations, important 3D FOV features, and lateral scan range constraints for this class of scanning arrangements.

\section*{Funding}
NIH R01MH117111; Arnold and Mabel Beckman Foundation (Beckman Young Investigator Award); Kinship Foundation (Searle Scholar Award); and Rita Allen Foundation (Rita Allen Scholar Award).

%\section*{Acknowledgments}
%Acknowledgments, if included, should appear at the end of the document.

%Added by MK
\bibliographystyle{unsrt}
\bibliography{main.bib}
%\printbibliography

\end{document}